# Observation of universal topological magnetoelectric switching in multiferroic GdMn$_2$O$_5$

Haowen Wang[1#], Fan Wang[2#], Ming Yang[1], Yuting Chang[1], Mengyi Shi[1], Liang Li[1], Jun-Ming Liu[3], Junfeng Wang[1*], Shuai Dong[2*], and Chengliang Lu[1*]

[1] *Wuhan National High Magnetic Field Center and School of Physics, Huazhong University of Science and Technology, Wuhan 430074, China*

[2] *Key Laboratory of Quantum Materials and Devices of Ministry of Education, School of Physics, Southeast University, Nanjing 211189, China*

[3] *Laboratory of Solid State Microstructures, Nanjing University, Nanjing 210093, China*

**[Abstract]** Topological magnetoelectricity was recently revealed as an emergent topic, which opens a unique route to precisely control magnetoelectric functionality. Here we report the synchronous magnetic-electric-cycle operation of topological magnetoelectric switching in GdMn$_2$O$_5$. Compared with pure magnetic-cycle operation, this topological winding can be accessed in a much broader parameter space, i.e. orientation of magnetic field is not limited to the magic angle and the effect can persist up to the Curie temperature. The fine tuning of free energy landscape is responsible to this topological behavior.

\# The authors contributed equally to this work.
\* Corresponding authors: J. F. W (jfwang@hust.edu.cn), S.D. (sdong@seu.edu.cn), C.L.L. (cllu@hust.edu.cn).

Topological physics has been one of most active fields of condensed matter physics, and a large array of emergent phenomena have so far been discovered, including topological insulators/semimetals/superconductors, as well as their associated quantum spin/anomalous Hall effect and Majorana fermion, etc [1-6]. In fact, as a concept of mathematics, topology can explicitly or implicitly dominate various physical behaviors, not limited to the electron/phonon/photon bands in the momentum space. The merge of topology and ferroic systems has already yielded a totally different story that real-space textures of magnetic and/or electric diploes can be topological, including skyrmions/merons/vortics with integer winding numbers [7-11].

Recently, another branch of topological physics has been revealed in some multiferroics, which exhibits topological winding behaviors for particular magnetoelectric (ME) processes. For example, for quadruple perovskite $TbMn_3Cr_4O_{12}$, a topological non-orientable Roman surface was proposed to describe the three-dimensional trajectory of magnetism-induced polarization ($P$) [12,13]. Another breakthrough was the ME switching in $GdMn_2O_5$ which generated a topological winding number for half Mn's spins in response to a magnetic-cycle [14]. Interestingly, this topologically protected ME process can be understood as a ME crank in the quantum level.

The topological ME switching is both fundamentally and practically significant, since it may allow robust and precise control of the ME functionality. Note that in multiferroics the competing magnetic interactions can generally lead to multiple energy valleys separated by small barriers in the free energy landscape, which brings challenges to the deterministic manipulation of ME states since system may fall into one of the valleys randomly, especially under the thermal fluctuation [15,16]. However, the topological ME switching that have been discovered up to now are commonly special and fragile. For example, the Roman surface topological ME switching of $TbMn_3Cr_4O_{12}$ relies on the rotation of magnetic field $H$ which drags the spins to evolve coherently [12], while in $GdMn_2O_5$ a slight deviation of $H$ from the "magic angle" (i.e. 2º), or slight higher temperature (i.e. $T$=5 K) will completely destroy this topological behavior [14]. Therefore, it is urgent to explore generalized approach to access the topological ME, such as broader working temperature window, no limitation of $H$ orientation, as well as applicable to other materials.

$GdMn_2O_5$ is chosen to address this issue, since it belongs to a typical family of multiferroic $RMn_2O_5$ ($R$: rare earth element or Bi) which host the exchange striction mechanism and thereby large magnetically driven $P$ [17-20]. Fig. 1(a) shows the crystalline structure of $RMn_2O_5$ [21]. The octahedra occupied by $Mn^{4+}$ and pyramids filled with $Mn^{3+}$ form pentagons in the *ab*-plane. With such geometry, the five relevant antiferromagnetic (AFM) interactions between neighboring Mn ions of a pentagon cannot be satisfied simultaneously, giving rise to strong magnetic frustration [22]. Nevertheless, two AFM chains can be developed along the *a*-axis below the Néel temperature $T_N$~33 K for $GdMn_2O_5$, and within each zigzag chain Mn-spins are coupled antiferromagnetically [23-25]. A spontaneous *P* is generated majorly along the *b*-axis due to the symmetric exchange striction, which is proportional to dot product of the AFM vectors of the two chains, i.e. $P$~$L_1 \cdot L_2$. Because of the competing interactions in $GdMn_2O_5$, at least five ferroelectric phases and diverse ME effects have been identified [26]. An interesting characteristic of $GdMn_2O_5$ is that its *P* can be reversed by an electric field *E* poling on the paramagnetic phase [27]. If *H* is applied parallel to $L_1$ or $L_2$, the energy landscape at low-*T* can be modified specifically that some energy valleys with seriously flattened barriers are distinguished from the others. As a consequence, the spin texture of $GdMn_2O_5$ evolves following a certain trace, and the ME switching path leads to a full-circle spin rotation of one AFM chain, which can be defined by a topological winding number *Q*=1, sketched in Figs. 1(b-c) [14]. A signature feature of the model is that states 1 and 3 (2 and 4) are associated by space inversion symmetry. Therefore, it is natural to expect to utilize *E* to engineer the potential energy surface, and thus generate unusual ME switching similar to the sole *H*-control. Moreover, the *E*-engineering would lift the degeneration of related ME states, and thus *H* may not have to be precisely aligned and *T* could be higher without losing the delicate asymmetric barrier associated with Gd-spins, that is, a broader operation window is obtained.

With this motivation, here we incorporate *E* to switch the ME states. With the *H-E* cycle, the energy surface can be tuned as desired and the topological spin winding in $GdMn_2O_5$ can be obtained in a much broader (*H*, *T*) parameter space. In particular, the *H*-orientation is not limited to the "magic angle" and the ME switching can persist up to the ferroelectric Curie temperature $T_C$~33 K. More $RMn_2O_5$ with other *R* ions are also investigated for comparison.

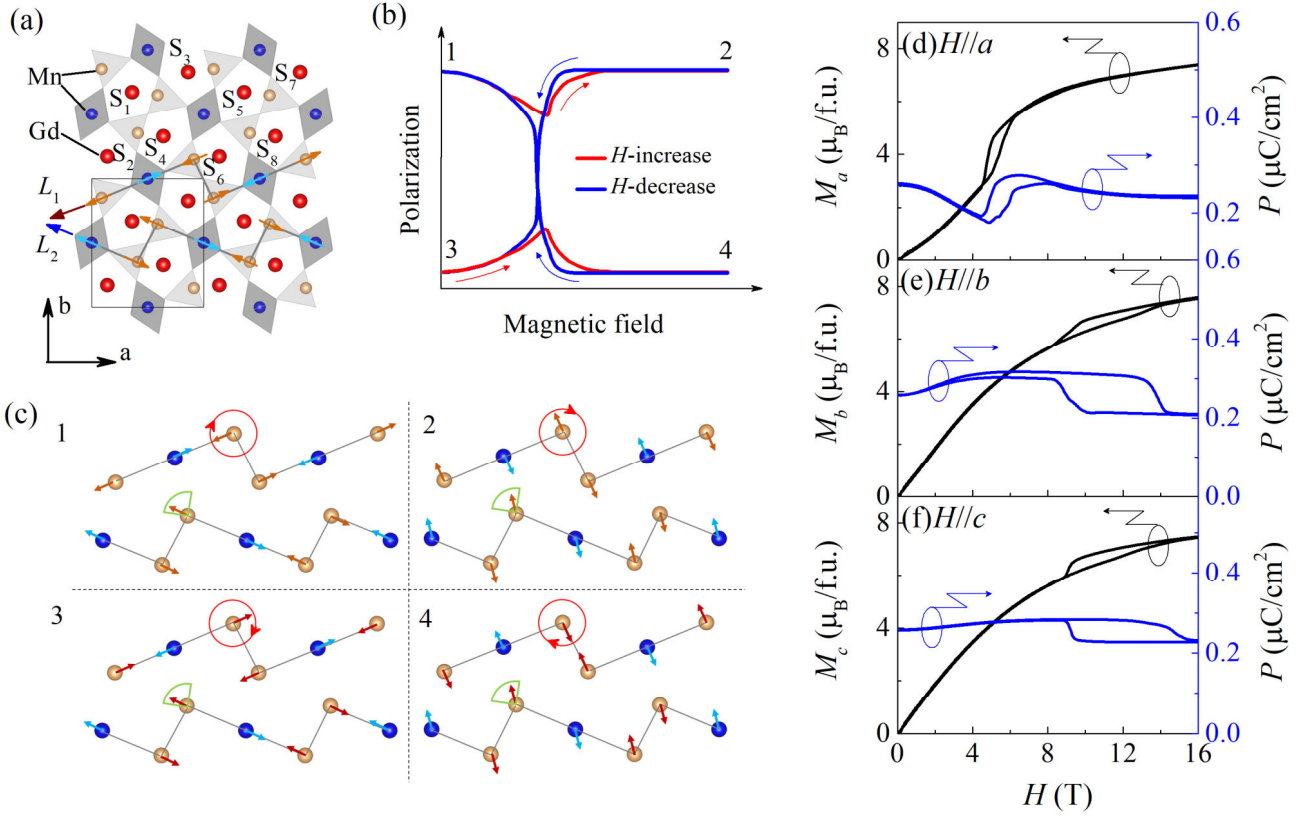

**Figure 1.** (a) Crystalline and magnetic structures of GdMn$_2$O$_5$. The antiferromagnetic chains with vectors $L_1$ and $L_2$, and Gd-spins $S_1$-$S_8$ have been labeled. (b) Schematic of topologically protected four-state ME-switching driven by $H$ applied at a "magic angle". The sequence 1-2-3-4 denotes the evolution of ME states during the $H$-sweeps. (c) The antiferromagnetic configurations of four states, in which the unidirectional full circle rotation of $L_1$ is indicated. (b-c) are plotted according to the scenario in Ref. [14]. (d-f) $H$-dependence of magnetization and polarization measured at 4.2 K with $H$ applied along the $a$-, $b$-, and $c$-axes, respectively.

Upon cooling the GdMn$_2$O$_5$ single crystal, a series of phase transitions are identified, including $T_{N1}$=40 K due to the onset of an incommensurate AFM order of Mn, $T_{N2}$=33 K arising from an incommensurate-commensurate transition which is accompanied by the emergence of spontaneous $P$ along the $b$-axis (i.e. $T_C$ for ferroelectricity), and $T_{Gd}$=12 K related to the magnetic ordering of Gd$^{3+}$, as shown in Fig. S3 in Supplementary Materials (SM) [28]. Gd$^{3+}$'s spin is nearly isotropic due to its $4f^7$ orbital filling, and its direction is generally determined by Mn's spin to fulfill the AFM interaction, which enhances the total polarization of GdMn$_2$O$_5$ remarkably to $P$~0.5 μC/cm$^2$ [23]. Spin-flop transition can be induced by $H$ applied along all three crystallographic directions, evidenced by the hysteresis loops in both magnetization ($M$) and $P$, shown in Fig. 1(d-f).

For the ME measurements, $P$ of the $b$-axis under various $H$ and $E$ were measured, and $E$ was always applied along the polar $b$-axis. A pulsed high magnetic field apparatus with pulse duration time $t\sim 10$ ms was used for the measurements. Before the measurements, the sample was cooled from $T>100$ K down to 4.2 K with $E=+10$ kV/cm and $H=0$, in order to fully align the ferroelectric domains. The obtained initial state, i.e. the state 1, has polarization $P_1\sim 0.26$ μC/cm², in good agreement with previous works [26,27]. In subsequent ME measurements, $H$ is applied along the $a$-axis. Fig. 2 shows $H$ dependence of $P$ measured under $\pm E$ at 4.2 K. With fixed $E=+10$ kV/cm, the cycle of $H$, i.e. sweeping up and down, drives the system to state 2 and then back to the original state 1, shown as the dashed curve in Fig. 2(a). This reversible process is topological trivial. The striking kink anomaly at $H_c\sim 5$ T in $P(H)$ is caused by the spin-flop in accordance with the $M(H)$ results shown in Fig. 1(d). The spin-flop transition remains visible close to $T_{N2}=33$ K (Fig. S5 in SM [28]), suggesting that it is associated with the AFM chains of Mn instead of the Gd's spin order below $T_{Gd}\sim 12$ K.

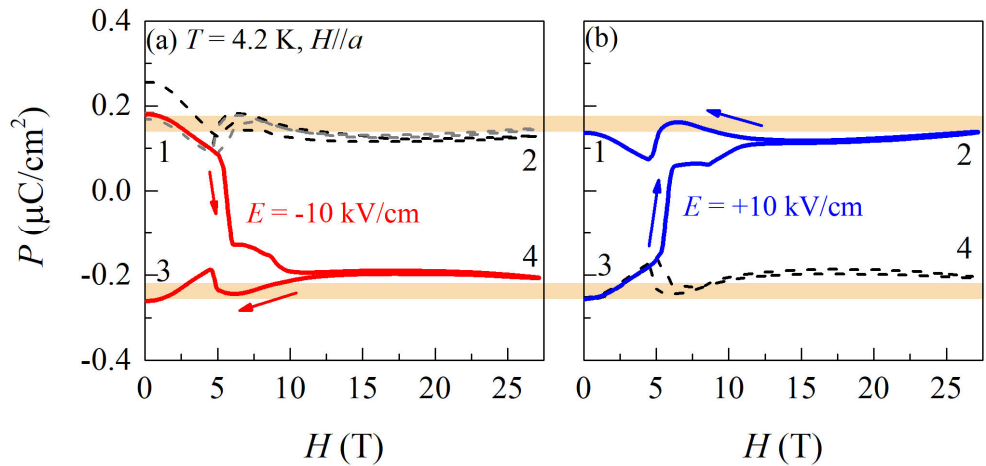

**Figure 2.** $H$ dependence of $P$ along the $b$-axis measured at 4.2 K with $H//a$. (a) ME switching upon the $H$-sweep. Dashed curve: reversible switching from state 1 to 2 and back to 1, under $E=+10$ kV/cm. Solid curve: irreversible switching from state 1 to 4 and then to 3 under $E=-10$ kV/cm. (b) The following $H$-$E$ cycle starting from state 3. Dashed curve: to state 4 and back to state 3 under $E=-10$ kV/cm. Solid curve: to state 2 and finally back to 1 under $E=+10$ kV/cm.

More interesting behavior is that using a negative $E=-10$ kV/cm, the $H$-sweep drive the system from state 1 to state 4 and then to state 3, leading to a striking hysteresis with sign-switching of $P$ (red curve in Fig. 2(a)). If $E$ is set back to +10 kV/cm subsequently in the second cycle, the system undergoes states 3, 2, and eventually return to the original state 1 (blue curve in Fig. 2(b)), when $H$ is swept up and then back to zero. In short, upon these two $H$-$E$ cycles, the four ME states are unidirectionally ergodic in a sequence of 1-4-3-2-1, where the spin winding direction (e.g. counterclockwise rotation of $L_1$) is in opposite to the 1-2-3-4-1 one in Ref. [14]. The introduction of $E$ during the $H$ sweeps makes the topological spin winding controllable.

The trajectory of 1-2-3-4-1 can also be obtained by reversing $E$ properly during the $H$-sweeps. As shown in Fig. 3(a), starting from state 1, $H$-sweep with $E$=+10 kV/cm drives the system to state 2. Subsequently, a negative electric field $E$=-10 kV/cm is applied at $H_{max}$, and thus the system is moved to state 3 as $H$ is ramped down to zero. A symmetric sequence of 3-4-1 can be accomplished by reversing $E$ at $H_{max}$ when further $H$-sweep is carried out, shown in Fig. 3(b). As a consequence, the 1-2-3-4-1 switching behavior is realized, where the spin winding direction is the same as the one in Ref. [14]. These ME switches reveal the efficient role of $E$ in tuning the energy barriers of various ME states.

Another prominent feature of ME switches is that the remarkable divergence of $P(H)$ only appears when $H$>$H_c$, despite the on/off or sign of $E$. Therefore the spin-flop transition at $H_c$ is the key point in tuning the unusual ME switching. To further confirm this point, a series of $P(H)$ curves were measured with different maximum field $H_{max}$ (Fig. S6 in SM [28]). It is found that after a cycle the $P(H=0)$ is different from the original one only when $H_{max}$>$H_c$. In Fig. 3(c), a nonzero $\Delta P$=$P_1$-$P_3$, i.e. difference in polarization between states 1 ($P_1$) and 3 ($P_3$), emerges right at $H_c$~5 T, and gets saturated above ~10 T, evidencing the intimate correlation between the four-states ME switching and the spin-flop transition. As mentioned above, the spin-flop transition is due to collective rotation of Mn-spins, which remains discernible up to $T_{N2}$=33 K. As a consequence, $\Delta P$ remains detectable as $T$ is close to $T_{N2}$, shown in Fig. 3(d) (Fig. S7 in SM [28]). Here $\Delta P$ is as large as ~0.5 µC/cm$^2$ at $T$=1.7 K, which is in good agreement with previous work [23].

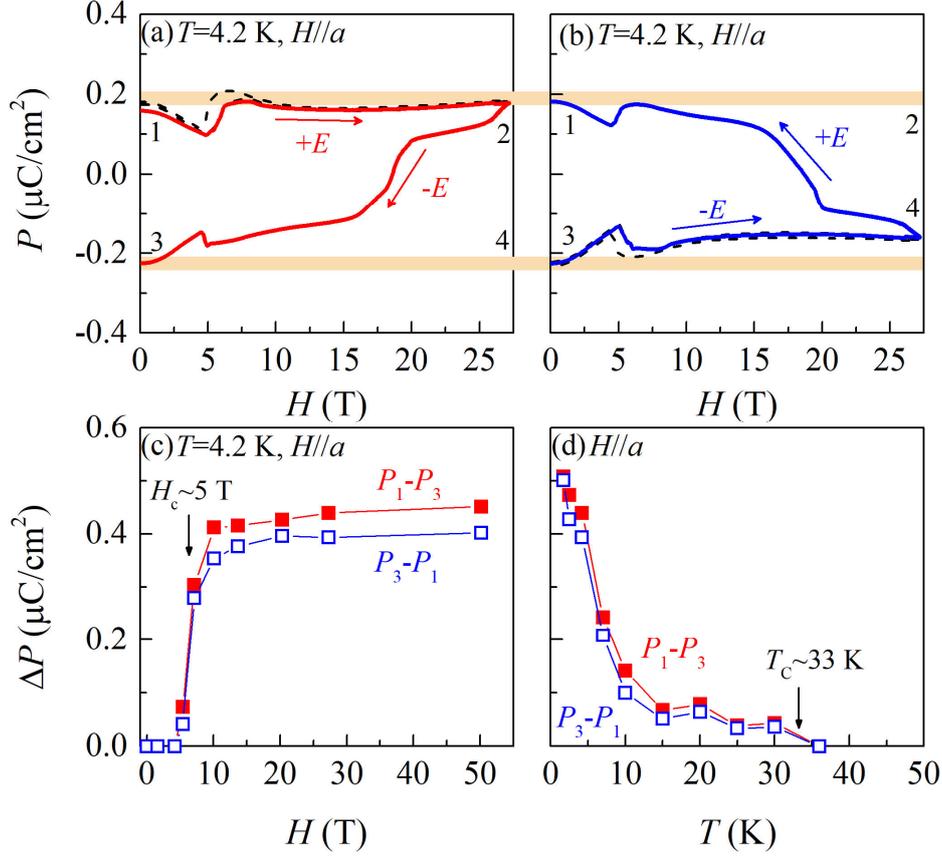

**Figure 3.** $H$ dependence of $P$ measured with $H//a$ under various $E=\pm10$ kV/cm. (a) The cycle starting from state 1. $E$ is reversed at $H_{max}$. (b) The following cycle starting from state 3. $E$ is reversed at $H_{max}$. (c) $\Delta P$ after a cycle of $H$-sweeps with different $H_{max}$ at 4.2 K. Nonzero values appear when $H>H_c\sim 5$ T. (d) $\Delta P$ as a function of temperature, which persists up to $T_{N2}=T_C\sim 33$ K. In (c) and (d), the data in red and blue are derived from $P(H)$ measured under $E=-10$ kV/cm and $E=+10$ kV/cm, respectively.

As discussed above, the spin-flop transition is crucial to create the four-states ME switching, and similar transition associated with hysteresis loops in both $M(H)$ and $P(H)$ are also observed for cases of $H//b$-axis and $H//c$-axis (Fig. 1(e-f)). In this sense, it would be possible to access the four-states ME switching with $H$ applied along other directions. To verify this point, extensive ME measurements were performed under various $E$ and $H$ (More data are shown in Fig. S8 in SM [28]). Before the measurements at every selected $\theta$ (relative angle between $H$ and the $a$-axis), the system was uniformly preset to state 1. Then, $P(H, \theta)$ curves were measured under $E=\pm10$kV/cm. Indeed, with the assist of $E$, the four-states ME switching is commonly obtained, despite the direction of $H$ in the $ab$-plane, as shown in Figs. 4(a-d).

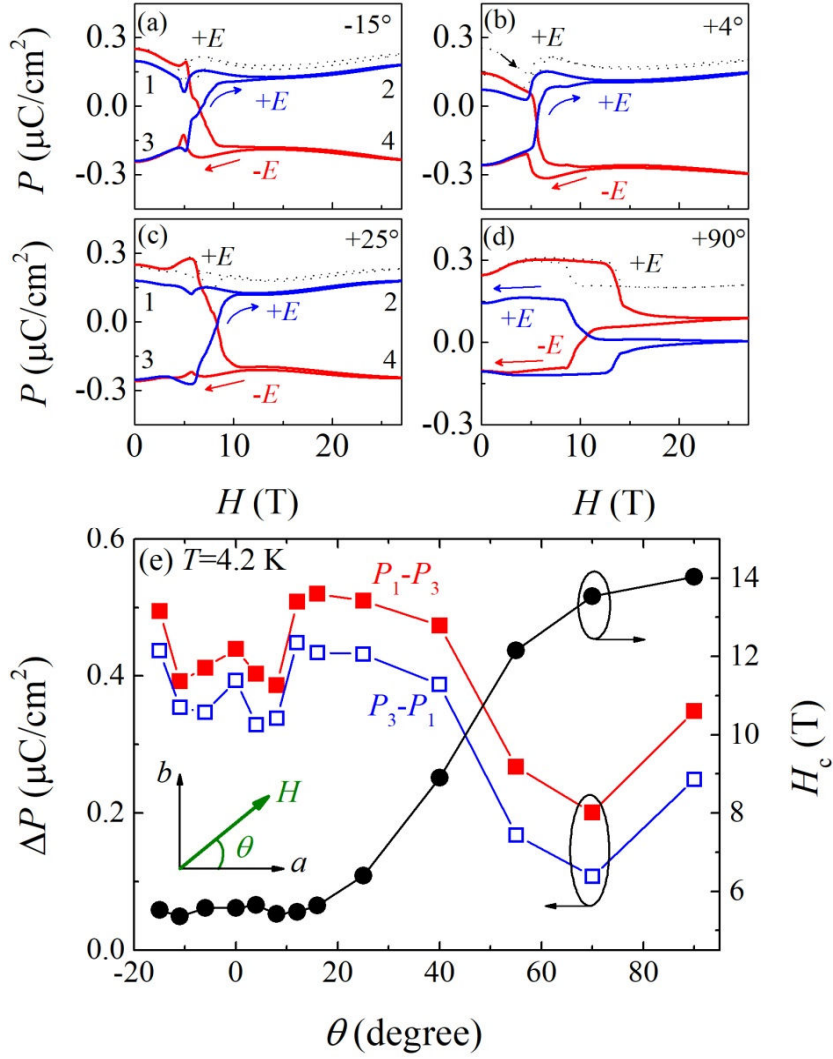

**Figure 4.** The angle dependence of ME switching. $P$ is measured under various $E= \pm 10$ kV/cm at $T$=4.2 K. During the measurements, $H$-sweeps were applied along various directions in the $ab$-plane: (a) $\theta$=-15°, (b) $\theta$=+4°, (c) $\theta$=+25°, and (d) $\theta$=+90°. (e) $\Delta P$ and the critical field $H_c$ as a function of $\theta$ at 4.2 K. The inset of (e) shows geometry of the measurements. For $\Delta P(\theta)$ in (e), the data in red and blue are derived from $P(H)$ measured under $E$=-10 kV/cm and $E$=+10 kV/cm, respectively.

While the four-states ME switching is qualitatively robust, the magnitude of the hysteresis, i.e. $\Delta P=P_1-P_3$, shows dependence on the direction of $H$, as plotted in Fig. 4(e). $\Delta P(\theta)$ is symmetrically distributed around $\theta$=0°, while clear anomaly of $\Delta P$ can be identified when $H$ is near the magic angle, i.e. $\theta=\pm 10°$, evidencing the contribution of $H$-orientation in flattening the energy barriers. Nevertheless, we didn't obtain the perfect 4-states switching behavior given of Ponet *et al.* [14]. This is probably because that the 4-states switching relying solely on the magic angle is extremely sensitive to the $H$-direction, sample difference, and other extrinsic perturbations. Similar to the case of $H//a$-axis, the large ME hysteresis always emerges once $H_{max}>H_c$, confirming the crucial role of the spin-flop transition in inducing the four-states switching. The derived $H_c$ as a function of $\theta$ is also shown in Fig. 4(e). Obviously, $H_c$ remains ~5 T as $H$ is near the $a$-axis, indicating the dominant role of the spin-flop transition driven by $H//a$ in tuning the four-states switching. For $\theta>20°$, $H_c$ increases with $\theta$ quickly, and eventually reaches $H_c$~14 T at $\theta=90°$ ($H//b$-axis). This coincides with the fast variation in $\Delta P(\theta)$. In comparison with $H//b$-axis, the application of $H//a$-axis is more efficient in generating a remarkable ME hysteresis. Similar to the cases of $H//a$ and $H//b$, the application of $H//c$ can also induce large ME hysteresis with the assist of $E$ operation. Since a much higher $H_c$ is required to trigger the spin-flop transition for $H//c$ (Fig. 1(f)), a larger $E$ would be needed to generate an ideal 4-states switching, which is responsible for the absence of sign switching of $P$ under $H//c$ and $E$=10 kV/cm in the present work. The data for $H$ applied along different crystalline axes are presented in SM (Fig. S9) [28].

To understand the $E$-assisted universal topological ME switching, a model simulation based on the Hamiltonian proposed in Ref. [14] has been performed, which can reproduce the experimental observations successfully. First, we replicated the result of Ref. 14, namely the process without $E$. When $H$ is applied along the so-called magic angle (e.g. 10° from the $a$-axis), the topological cycle indeed appears due to the finely tailored free energy landscape (Figs. S10 in SM [28]). In contrast, when $H$ is applied along the $a$-axis (i.e. non-magic angle), the free energy landscapes are tuned, and then the transition path will become trivially reversible, as shown in Figs. 5(a-b). Second, to incorporate the electric field, an additional electrostatic energy term is introduced to the Hamiltonian [28], which indeed restores to the topological cycle of $P_b$ via the -$E/E$ protocol, as shown in Fig. 5(a). The corresponding free energy landscapes are shown in Figs. 5(c-d). Clearly, our simulation successfully reproduces the experimental results with/without $E$, and the underlying mechanism is the fine tuning of free energy landscape.

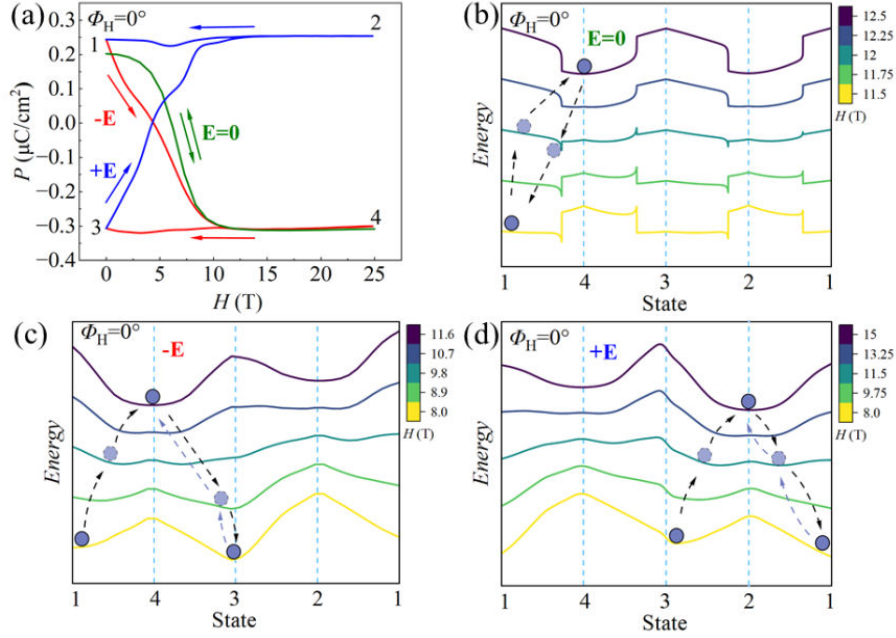

**Figure 5.** Simulation of the magnetoelectric switching. (a) The cycle of $P_b$ for $H//a$, with/without $E$. (b-d) The free energy landscapes for $H//a$. (b) Without $E$. The transition path during the magnetic cycle becomes reversible. (c) With a negative $E$. The $-P$ states (3 & 4) are more preferred slightly. (d) With a positive $E$. The $+P$ states (1 & 2) are more preferred slightly. The paths are determined by the steepest descent method in free energy, instead of the known transition end states.

Although Ref. [14] claimed the crucial role of 4$f$ magnetic ion for such a topological magnetoelectricity, there were no direct experimental verification. Here we prepare two additional groups of materials: 1) $DyMn_2O_5$ and $ErMn_2O_5$ with large $f$-electron magnetic moments of $R$ ions [20, 31], 2) $BiMn_2O_5$ and $YMn_2O_5$ with non-magnetic $R^{3+}$ ions [18, 32]. The ME measurements indeed confirm that the four-states ME switching is observed in $DyMn_2O_5$ and $ErMn_2O_5$, similar to $GdMn_2O_5$. However, there is no $P(H)$ hysteresis at $H=0$ for both $BiMn_2O_5$ and $YMn_2O_5$, that is, $P$ goes back to the starting point after $H$-cycle, no matter with or without the application of $E$.

Additional model simulations are also performed to understand the effect coming from the $f$-magnetic moment. Without the $f$-moment, the 3$d$-spin only model is trivially simple, which can not lead to the topological winding number during the magnetic cycle. The key to achieve the topological winding number during the switching cycle is to enable one antiferromagnetic Mn chain to flop, which needs to cross a potential barrier of magnetocrystalline anisotropy. In Ref. [14], a magnetic field along the magic angle is used to flop the 4$f$ moment first at high fields, and then the exchange interaction between $Mn^{3+}$ and $Gd^{3+}$ assists the adjacent antiferromagnetic Mn chain to complete the flop. In our study, the electric field can help the flop beyond the magic angle, via the electrostatic energy. However, without the 4$f$ magnetism, the potential barrier is too high to be climbed over under a moderate electric field, as observed in our simulation and experiment (shown in Figs. 11 and 12 in SM [28]).

In summary, our work has illustrated a generalized route to access the topologically protected ME switching in multiferroic $GdMn_2O_5$, which essentially depends on the spin-flop transition of Mn rather than the canting angle of AFM chains. The electric field $E$ is crucial to determine the final ME state. Our model simulations can reproduce the experimental findings, and reveal the crucial role of electric field in tuning the free energy landscape. In addition, it is also found that the 4$f$ moment is essential in flattening the energy barriers.

**Acknowledgements.** This work is supported by the National Natural Science Foundation of China (Grant Nos. 12174128, 12104351, 11834002, 92163210, and 12074135), Hubei Province Natural Science Foundation of China (Grant No. 2021CFB027), China Postdoctoral Science Foundation (Grant No. 2023M731209), and the Interdisciplinary program of Wuhan National High Magnetic Field Center (WHMFC) at Huazhong University of Science and Technology (Grant No. WHMFC202205).

**References**:

[1] M. Z. Hasan and C. L. Kane, Rev. Mod. Phys. **82**, 3045 (2010).

[2] X.-L. Qi and S.-C. Zhang, Rev. Mod. Phys. **83**, 1057 (2011).

[3] X. Wan, A. M. Turner, A. Vishwanath, and S. Y. Savrasov, Phys. Rev. B **83**, 205101 (2011).

[4] M. N. Ali, J. Xiong, S. Flynn, J. Tao, Q. D. Gibson, L. M. Schoo, T. Liang, N. Haldolaarachchige, M. Hirschberger, N. P. Ong, and R. J. Cava, Nature **514**, 205 (2014).

[5] K. Wang, D. Graf, H. Lei, S. W. Tozer, and C. Petrovic, Phys. Rev. B **84**, 220401 (2011).

[6] J. Sau, S. Simon, S. Vishveshwara, and J. R. Williams, Nat. Rev. Phys. **2**, 667 (2020).

[7] T. Choi, Y. Horibe, H. T. Yi, Y. J. Choi, W. Wu, and S. W. Cheong, Nat. Mater. **9**, 253 (2010).

[8] X. Z. Yu, S. Seki, S. Ishiwata, and Y. Tokura, Science **336**, 198 (2012).

[9] F.-T. Huang and S.-W. Cheong, Nat. Rev. Mater. **2**, 1 (2017).

[10] Y. Tokura and N. Nagaosa, Nat. Commun. **9**, 3740 (2018).

[11] K. Du, B. Gao, Y. Wang, X. Xu, J. Kim, R. Hu, F.-T. Huang, and S.-W. Cheong, npj Quantum Mater. **3**, 33 (2018).

[12] G. X. Liu, M. C. Pi, L. Zhou, *et al.*, Nat. Commun. **13**, 2373 (2022).

[13] Z. Wang, Y. Chai, and S. Dong, Phys. Rev. B **108**, L060407 (2023).

[14] L. Ponet, S. Artyukhin, Th. Kain, J. Wettstein, A. Pimenov, A. Shuvaev, X. Wang, S.-W. Cheong, M. Mostovoy, and A. Pinmenov, Nature **607**, 81 (2022).

[15] S.-W. Cheong, D. Talbayev, V. Kiryukhin, and A. Saxena, npj Quantum Mater. **3**, 19 (2018).

[16] S. Dong, J.-M. Liu, S.-W. Cheong, and Z. Ren, Adv. Phys. **64**, 519 (2015).

[17] N. Hur, S. Park, P. A. Sharma, J. S. Ahn, S. Guha, and S.-W. Cheong, Nature **429**, 392 (2004).

[18] J. W. Kim, S. Y. Haam, Y. S. Oh *et al.*, Proc. Natl. Acad. Sci. **106**, 15573 (2009).

[19] D. Higashiyama, S. Miyasaka, N. Kida, T. Arima, and Y. Tokura, Phys. Rev. B **70**, 174405 (2004).

[20] N. Hur, S. Park, P. A. Sharma, S. Guha, and S. W. Cheong, Phys. Rev. Lett. **93**, 107207 (2004).

[21] S. C. Abrahams and J. L. Bernstein, J Chem. Phys. **46**, 3776 (1967).

[22] G. R. Blake, L. C. Chapon, P. G. Radaelli, S. Park, N. Hur, S. W. Cheong, and J. Rodríguez-Carvajal, Phys. Rev. B **71**, 214402 (2005).

[23] N. Lee, C. Vecchini, Y. J. Choi, L. C. Chapon, A. Bombardi, P. G. Radaelli, and S. W. Cheong,

Phys. Rev. Lett. **110**, 137203 (2013).

[24] A. Muñoz, J. A. Alonso, M. T. Casais, M. J. Martínez-Lope, J. L. Martínez, and M. T. Fernández-Díaz, Phys. Rev. B **65**, 144423 (2002).

[25] C. Vecchini, L. C. Chapon, P. J. Brown, T. Chatterji, S. Park, S. W. Cheong, and P. G. Radaelli, Phys. Rev. B **77**, 134434 (2008).

[26] S. H. Bukhari, Th. Kain, M. Schiebl, A. Shuvaev, A. Pimenov, A. M. Kuzmenko, X. Wang, S. W. Cheong, J. Ahmad, and A. Pimenov, Phys. Rev. B **94**, 174446 (2016).

[27] S. H. Zheng, J. J. Gong, Y. Q. Li *et al.*, J Appl. Phys. **126**, 174104 (2019).

[28] See Supplemental Material at ** , which includes Refs. [29, 30], for more theoretical and experimental results.

[29] H. Kimura, S. Kobayashi, Y. Fukuda, T. Osawa, Y. Kamada, Y. Noda, I. Kagomiya, and K. Kohn, J. Phys. Soc. Jpn. **76**, 074706 (2007).

[30] G. E. Johnstone, R. A. Ewings, R. D. Johnson, C. Mazzoli, H. C. Walker, and A. T. Boothroyd, Phys. Rev. B **85**, 224403 (2012).

[31] D. Higashiyama, S. Miyasaka, and Y. Tokura, Phys. Rev. B **72**, 064421 (2005).

[32] R. P. Chaudhury, C. R. dela Cruz, B. Lorenz, Y. Sun, C.-W. Chu, S. Park, and S.-W. Cheong, Phys. Rev. B **77**, 220104 (2008).